\documentclass{aa}
\usepackage{graphicx,epsfig,txfonts,natbib}
\def\kms{\rm km \; s^{-1} }

\begin{document}

\title{Small-scale flows in SUMER and TRACE high-cadence co-observations}

\author{M.S. Madjarska\inst{1} \and J.G. Doyle\inst{2}}

\offprints{madjarska@mps.mpg.de}
\institute{Max-Planck-Institut f\"{u}r Sonnensystemforschung, Max-Planck-Str. 2,
37191 Katlenburg-Lindau, Germany
         \and	 
	 Armagh Observatory, College Hill, Armagh BT61 9DG, N. Ireland}

\date{Received date, accepted date}
	 
\abstract
{We report on the physical properties of small-scale transient flows observed 
simultaneously at high cadence with the SUMER spectrometer and the TRACE  imager 
in the plage area of an active region.}{Our major objective is to provide a better 
understanding of the nature of transient phenomena  in the solar atmosphere by 
using high-cadence imager and spectrometer co-observations at similar 
spatial and temporal resolution.}{A sequence of TRACE Fe~{\sc ix/x}~$\lambda$171~\AA\ 
and high-resolution MDI images were analysed together with simultaneously obtained SUMER 
observations in spectral lines covering a temperature range from 10\,000 K to 1 MK.}
{We reveal the existence of numerous transient flows in small-scale loops (up to 
30~Mm) observed in the plage area of an active region. These flows have 
temperatures from 10\,000~K (the low temperature limit of our observations) to
250\,000 K. The coronal response of 
these features is uncertain due to a blending of the observed coronal line
Mg~{\sc x}~$\lambda$624.85~\AA. The duration of the events ranges from 60~{\rm s} to 
19~{\rm min} depending on the loop size. Some of the flows reach  supersonic velocities.}
{The Doppler shifts often associated with explosive events or bi-directional jets can 
actually be identified with flows (some of them reaching supersonic velocities) in 
small-scale loops. Additionally, we demonstrate how a line-of-sight effect can 
give misleading information on the nature of the observed  phenomena if only 
either an imager or a spectrometer is used.}

\authorrunning{Madjarska \& Doyle}
\titlerunning{Small-scale flows in SUMER and TRACE}
	 
\maketitle

\begin{figure*}[ht!]
\centering
\vspace{6cm}
        \includegraphics{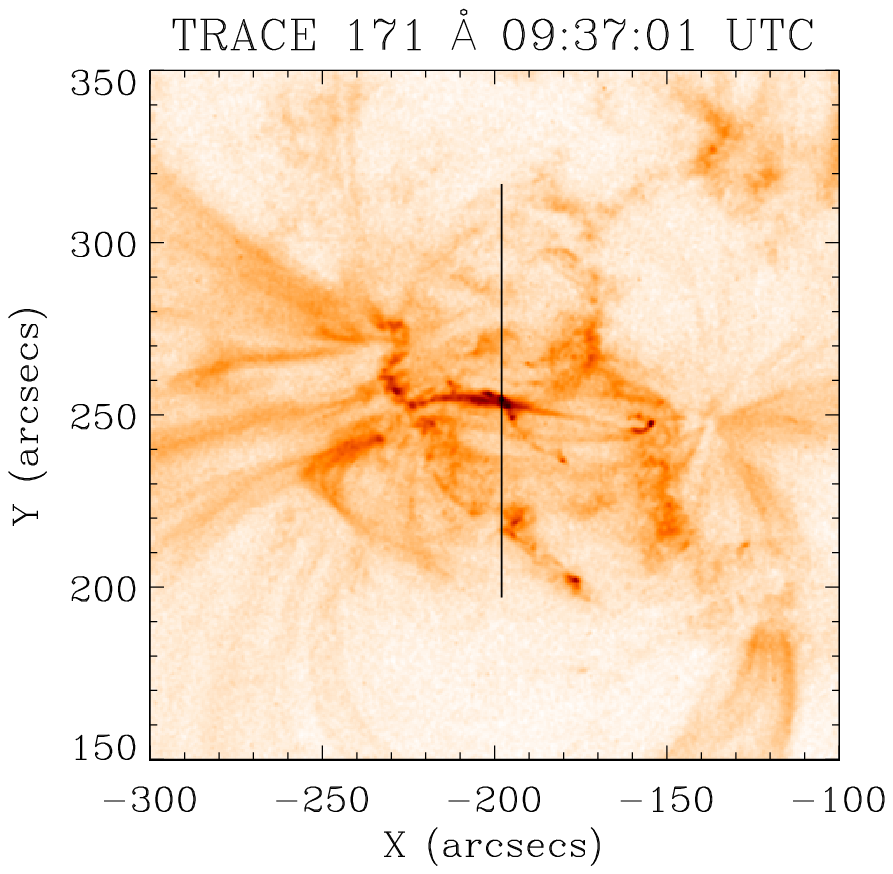}
	\includegraphics{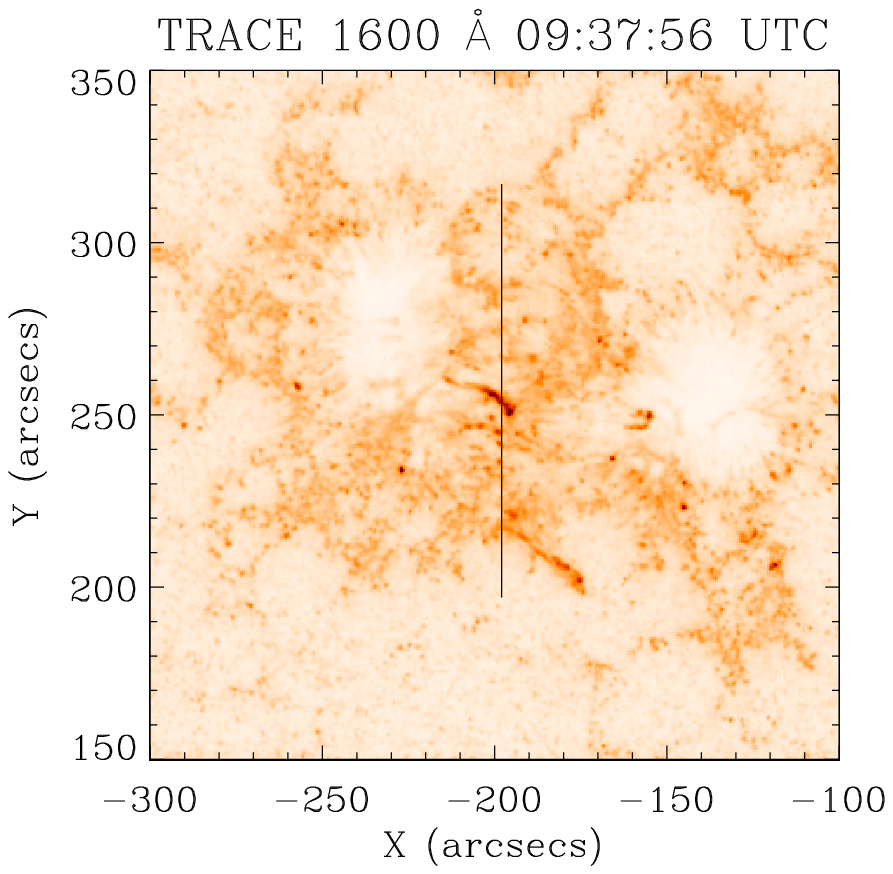}
	\includegraphics{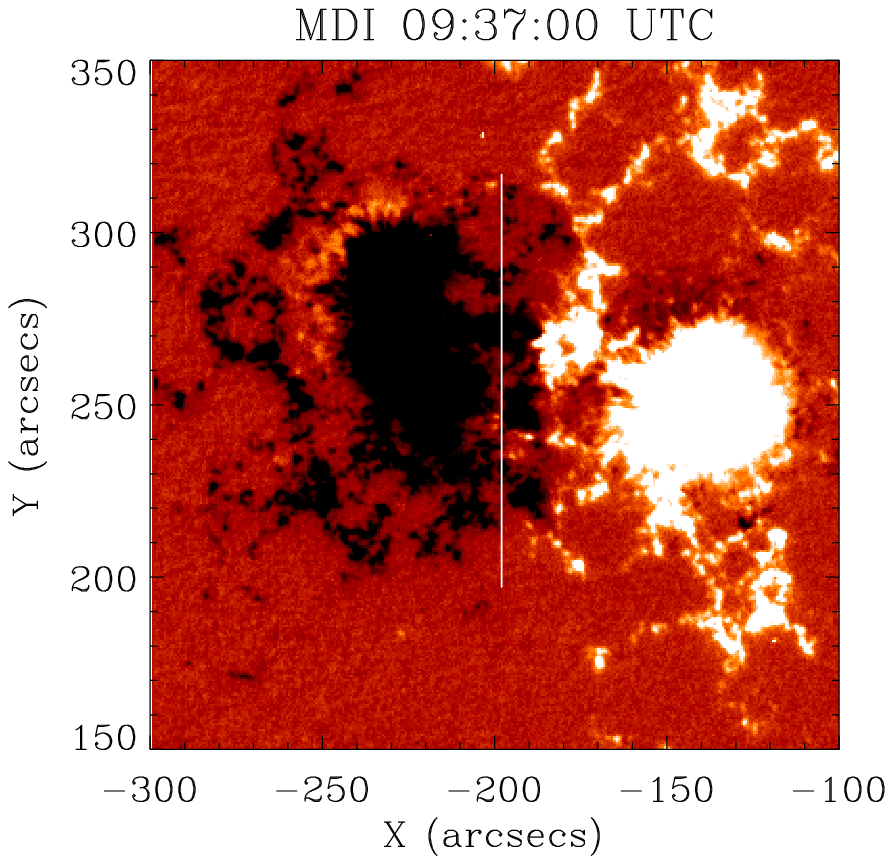}

	\caption{Colour table reversed TRACE~$\lambda$171~\AA\ (left) and 
	TRACE~$\lambda$1600~\AA\ (middle) images and MDI high-resolution 
	magnetogram scaled from -200 to 200 {\rm G} (right) of the active 
	region NOAA 08558 observed on 1999 June 2  with the SUMER slit 
	position over-plotted.}
	\label{fig1}
	\end{figure*}

\section{Introduction}

The solar atmosphere is highly dynamic on all scales seen both in spectroscopic and 
imager data. The dynamics is usually witnessed by non-Gaussian spectral line 
profiles, a strong radiance increase or proper motion of bright features (dark when 
seen in absorption). These events are commonly named transient phenomena due to 
their short duration. They have already been intensively studied for almost 
three decades during the High-Resolution Telescope and Spectrometer (HRTS), Yohkoh, 
Solar and Heliospheric Observatory (SoHO) and Transition Region And Coronal Explorer 
(TRACE)  missions and it is strongly believed that they may contribute to both the 
coronal heating and solar wind generation. However, it is often difficult or even 
impossible to derive their spatial scale and even to identify their true nature because 
of the limitations of the existing instruments. Either high-cadence spectroscopic 
rastering, faster than the lifetime of these phenomena, or lower cadence rastering and 
simultaneous imaging can provide their correct identification and better 
understanding of the physical mechanisms involved.

Transient flows were studied since the Skylab mission \citep{1992ApJ...401..754M} 
and later during the SoHO mission 
\citep[for a recent overview see][]{2006A&A...452.1075D}. Only recently a  
transient flow in a small-scale quiet-Sun loop was reported by \citet{2004A&A...427.1065T} 
detected in Solar Ultraviolet Measurement of Emitted Radiation (SUMER) data. The 
observations were taken in a very high-cadence rastering mode with 3~{\rm s} exposure 
time which permitted a `snapshot' of the area to be obtained. Transient flows in loops 
can be created by  heating or pressure imbalance between the footpoints of a loop 
or by asymmetry in the footpoint areas 
\citep{1982ApJ...258L..49B, 1991ApJ...382..338S, 1995A&A...294..861O, 1995A&A...300..549O}. 
The heat deposition is believed to be located at  the footpoints of the loop 
\citep{2006ApJ...642..579S}.

The phenomena which have all the characteristics of a transient feature are the 
so-called explosive events also known as bi-directional jets. They were seen both 
in the quiet Sun and active regions and  were first observed with HRTS by 
\citet{1983ApJ...272..329B} and later during 
the SoHO mission in SUMER observations \citep{1997Natur.386..811I}. They are identified 
by their non-Gaussian profiles and were  registered in spectral lines with formation 
temperatures from  4~10$^4$~K up to 6~10$^5$~K. No response was found so far at coronal 
temperatures \citep{2002A&A...392..309T}. Their spatial size estimated from the appearance 
along a spectrometer slit is 3\arcsec--5\arcsec. Their lifetime  ranges
from 60~{\rm s} to 300~{\rm s}. Explosive events are mostly known from their spectral 
characteristics. Observations showing them simultaneously  
 in imager and spectrometer data are very limited. \citet{2001ApJ...553L..81W} 
 used simultaneous TRACE and SUMER 
active region observations and found that short-term ($\le$ 5 {\rm min}) intensity 
fluctuations in TRACE~$\lambda$171~\AA\ data are on average 2.2 times larger 
in regions of reconnection than in a non-event region. The regions of reconnection 
were identified as the regions in which explosive event(s) were observed. The 
cadence of the TRACE observations was 50~{\rm s} which is far too low to 
identify any proper motion if present. \citet{2001A&A...378.1067I} made a detailed 
analysis of simultaneous SUMER Si~{\sc iv}~$\lambda$1393~\AA\ line profiles, 
TRACE~$\lambda$1550~\AA, $\lambda$1700~\AA\ and $\lambda$171~\AA\ passband images 
and Michelson Doppler Imager (MDI) high-resolution magnetograms. The author found 
that in most events the Si~{\sc iv}~$\lambda$1393~\AA\  line reveals plasma flows 
1--2 {\rm min} before the line core brightens suggesting that the plasma 
acceleration  precedes plasma compression and/or heating.

The large variety of transient phenomena  reported so far is mainly 
known from observations obtained  with one particular instrument and in a 
certain wavelength range. We aim at studying  in detail  small-scale transients 
and determine their plasma and spatial characteristics using simultaneous 
spectroscopy and vacuum ultra-violet imaging at the highest existing spatial, 
temporal and spectral resolution (in the spectral range in question). We will also 
demonstrate the importance of having both imager  as well as an imaging spectrometer 
data in order to avoid misinterpretation of an observed feature. In Sect.~2 we 
describe the observational material, its reduction  and the way the co-alignment 
between the different instruments was done. Sect.~3 presents the data analysis, 
the obtained results and their discussion. In Sect.~4 we discuss the importance 
of multi-instrument observations. Conclusions as well as the future perspectives on the 
subject are given in Sect.~5.

\section{Observations: SUMER, TRACE and MDI}

The events discussed here occurred in the plage area of the active region NOAA 8558 
on 1999 June 2 (Fig.~\ref{fig1}). Simultaneous SUMER, TRACE and MDI observations 
were taken during several hours (see below for more details). The  fields-of-views
(FOVs) of SUMER, TRACE and MDI are shown in Fig.~\ref{fig1}.

The SUMER spectrometer \citep{1995SoPh..162..189W,1997SoPh..170..105L} data (1.5\arcsec\  
spatial resolution) were taken on 1999 June 2 starting at 09:17~UTC and ending at 
11:02~UTC. A slit with a size of 0.3\arcsec\ $\times$ 120\arcsec\ was used exposing 
for 25~{\rm s} on detector B. The slit was pointed at the plage area of the active 
region between two sunspots of opposite polarities (Fig.~\ref{fig1}). Four 
spectral  windows were telemetered each with a size of 120 spatial $\times$ 50 
spectral pixels. The spectral line read-outs are shown in Table~\ref{table:1}. 
From all lines only  O~{\sc v}~$\lambda$629.73 \AA\ was taken on the bare
part of the detector.  At the start of the observations the 
spectrometer was pointed at solar disk coordinates  xcen~=~-217\arcsec\ (at 09:17~UTC) and 
ycen~=~257\arcsec. Subsequently, the observations were compensated for the solar rotation.
  The data were reduced using the standard procedures for flatfield, local gain and
 geometric distortion corrections.
The spectral analysis was made  with respect to a reference spectrum obtained by averaging 
over the entire dataset. We used the spectral atlas of \citet{2001A&A...375..591C},
\citet{1986ApJS...61..801S} and \citet{1987aisl.book.....K} to identify the spectral lines as 
well as cross-checking with the CHIANTI v5.2 database.


\begin{figure*}[ht!]
\centering
\vspace{20cm}

        \includegraphics{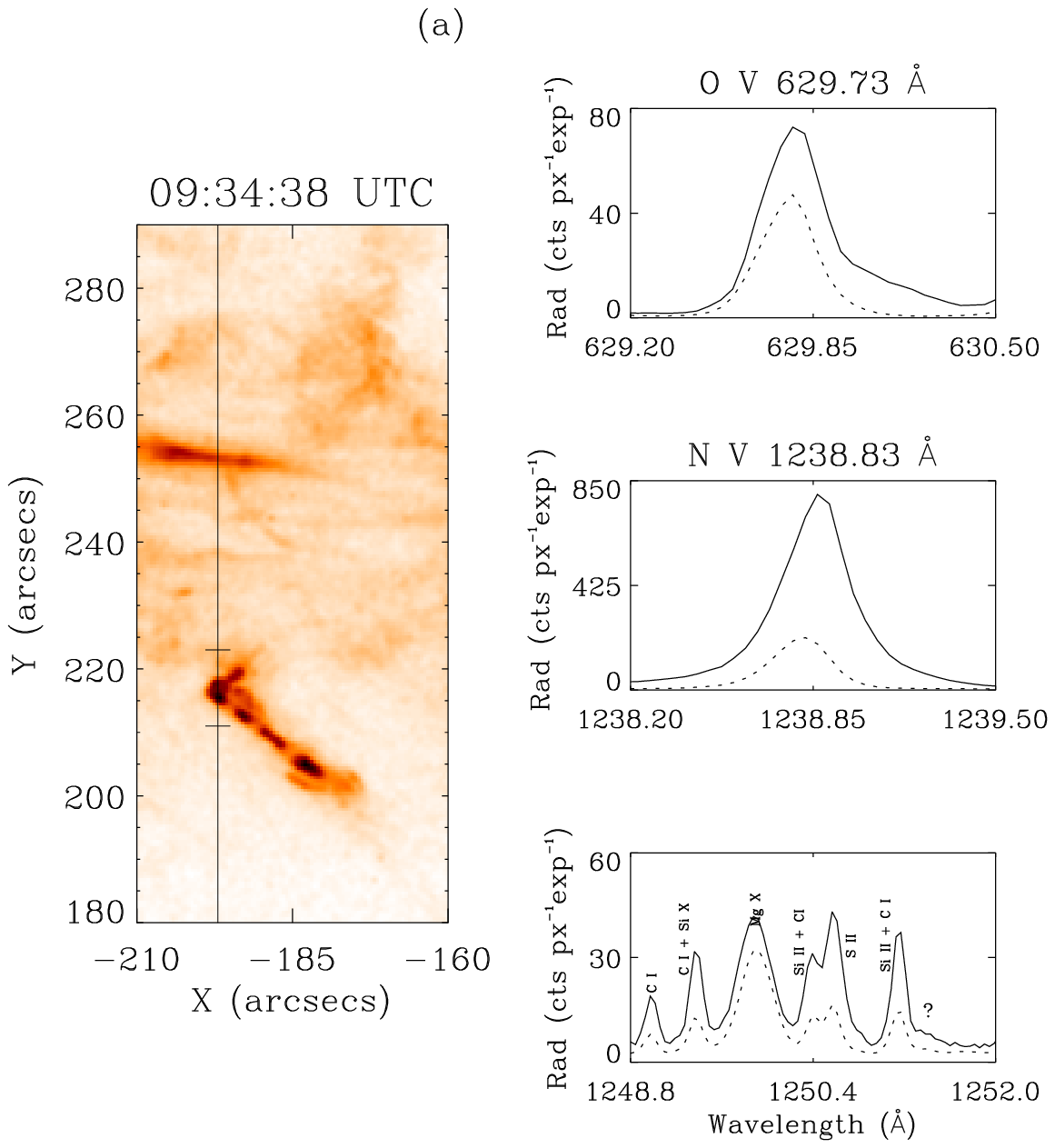}
	\includegraphics{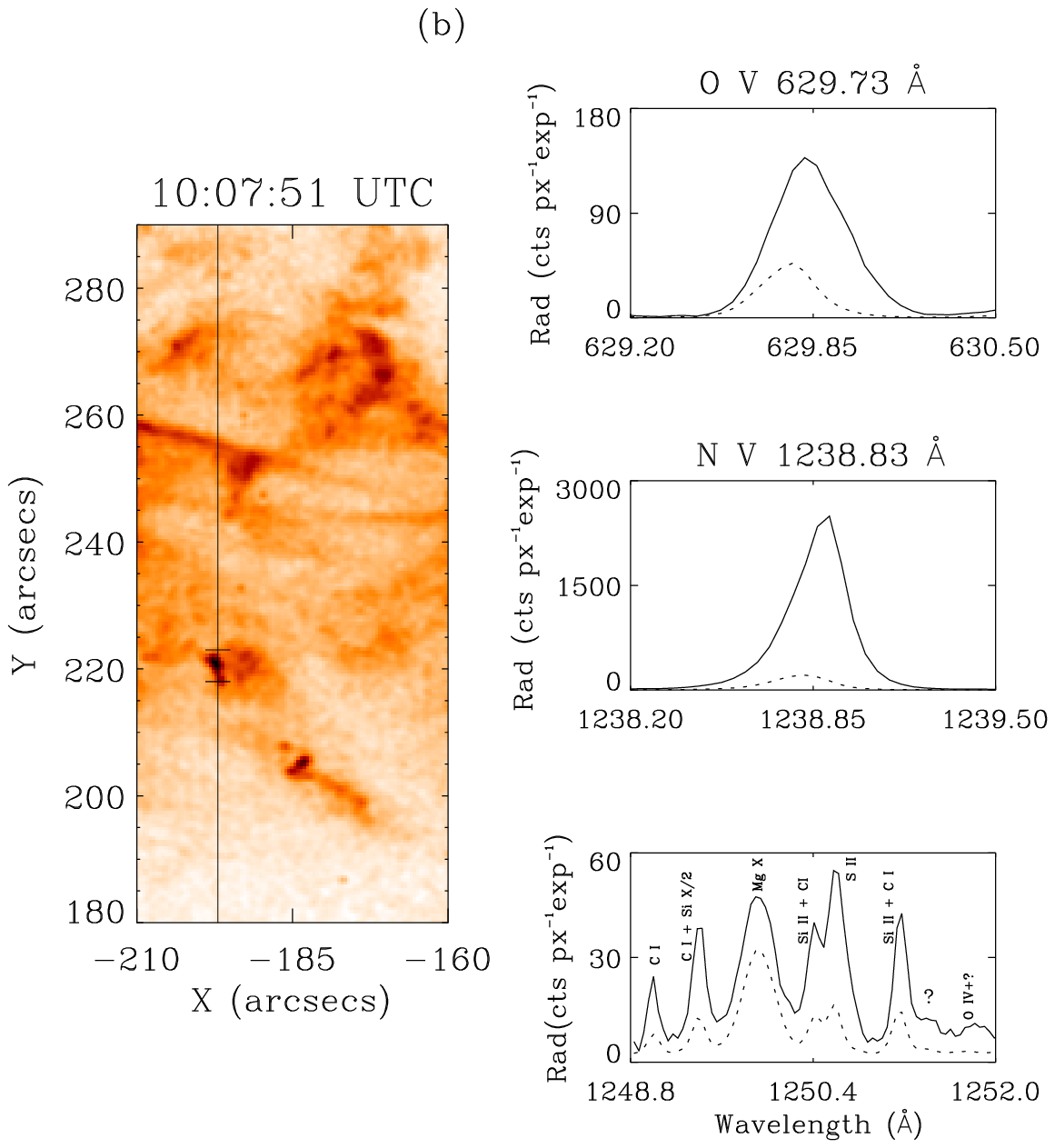}
	\includegraphics{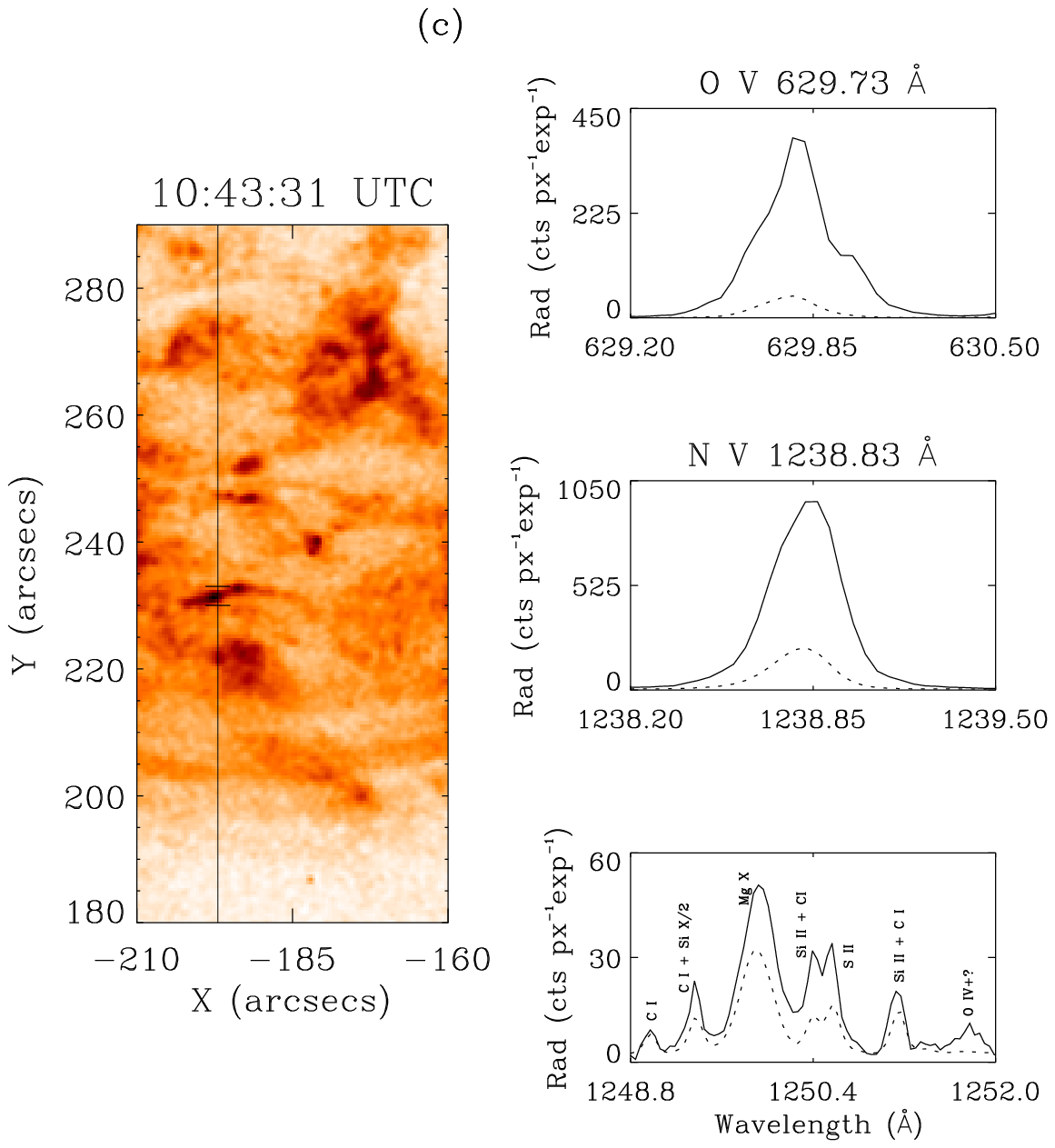}
	\includegraphics{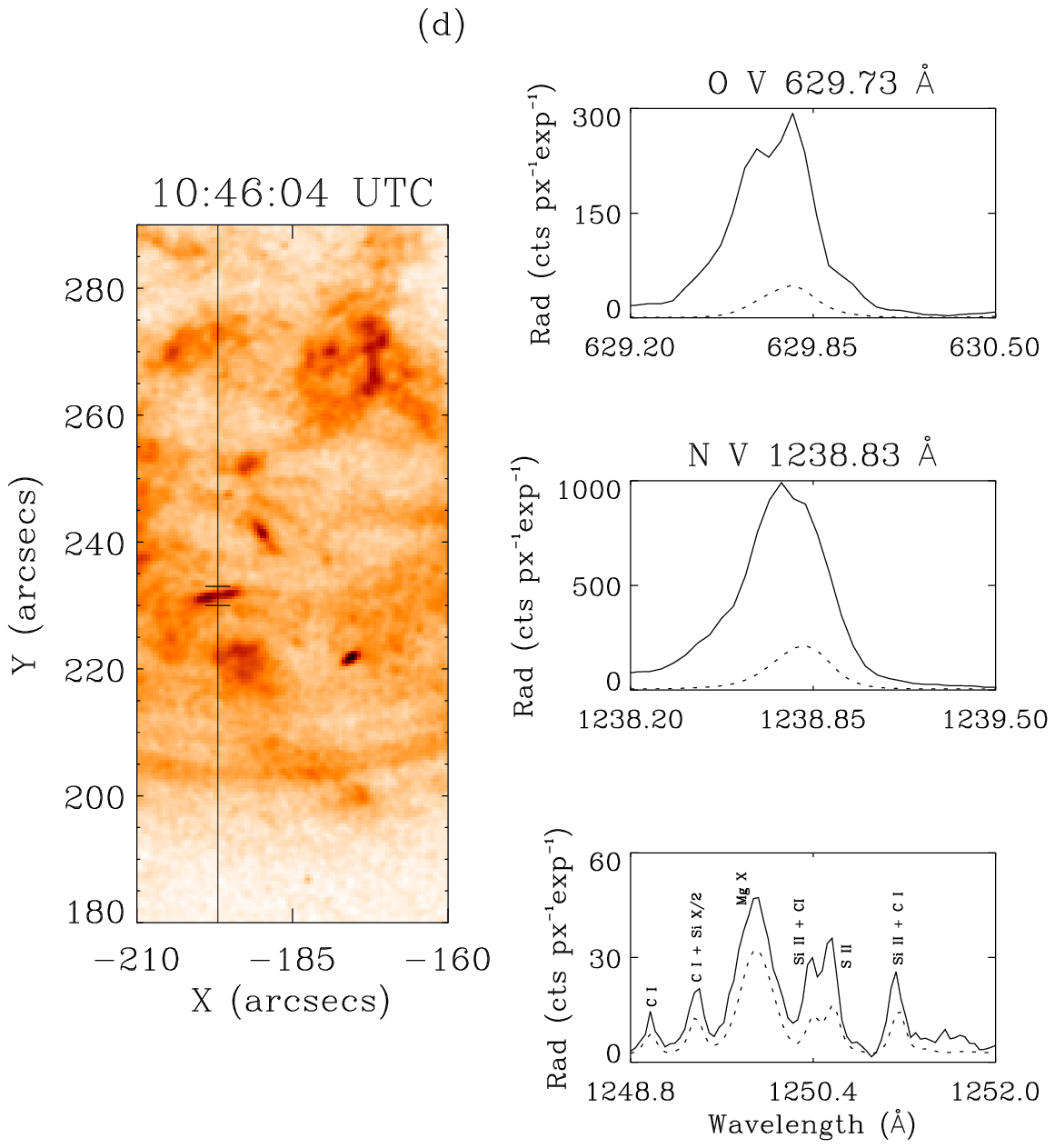}

	\caption{For all cases (a), (b), (c) and (d): {\bf left:} TRACE $\lambda$171 \AA\
	colour table reversed image showing the analysed feature. The vertical line 
	corresponds to the position of the SUMER slit, while the two horizontal lines 
	outline  the part of the slit  which was analysed in the SUMER data. {\bf right:} 
	Profiles of all observed spectral lines taken closest in time to the shown TRACE image.
	The dotted line profile corresponds to the reference spectrum.  An animation of the
	TRACE~$\lambda$171~\AA\ images can be seen online.}
	\label{fig2}
\end{figure*}

\begin{table}
\centering
\caption{The observed spectral lines. The expression `/2' means that the spectral
 line was observed in second order. The comment `blend' means that the spectral line is 
 blending a  close-by line.
\label{table:1}}

\begin{tabular}{c c c c c}
\hline\hline
Ion & $\lambda/$\AA & log(T)$_{max})$/K & Comment\\
\hline

N~{\sc v} & 1238.82 & 5.3 & \\
C~{\sc i} & 1248.00 & 4.0 & \\
& 1248.88 & & blend\\
C~{\sc i} & 1249.00 & 4.0 &  \\
O~{\sc iv}/2 & 1249.24 & 5.2&  blend\\ 
Si~{\sc x}/2 & 1249.40 & 6.1 & blend\\
C~{\sc i} & 1249.41 & 4.0 & \\     
Mg~{\sc x}/2 & 1249.90 & 6.1 & \\
O~{\sc iv}/2 &1250.25&5.2& blend\\
Si~{\sc ii} & 1250.09&4.1 &  \\
Si~{\sc ii} & 1250.41 &4.1 & \\
C~{\sc i} & 1250.42 & 4.0 & blend\\
S~{\sc ii} & 1250.58 &4.2 &\\
Si~{\sc ii} & 1251.16 &4.1 & \\
C~{\sc i} & 1251.17 &4.0 & blend\\
Si V&1251.39&5.5&?\\
O~{\sc iv}/2&1251.70&5.2&\\
& 1251.78 &&?\\
Si~{\sc i} & 1258.78 &4.1 &\\
S~{\sc ii} & 1259.53 &4.2 &blend\\
O~{\sc v}/2&1259.54&5.4&&  \\
Si~{\sc ii}&1260.44&4.1&\\
\hline
\end{tabular}
\end{table}


\begin{figure*}[ht!]
\centering
\vspace{12cm}

	\includegraphics{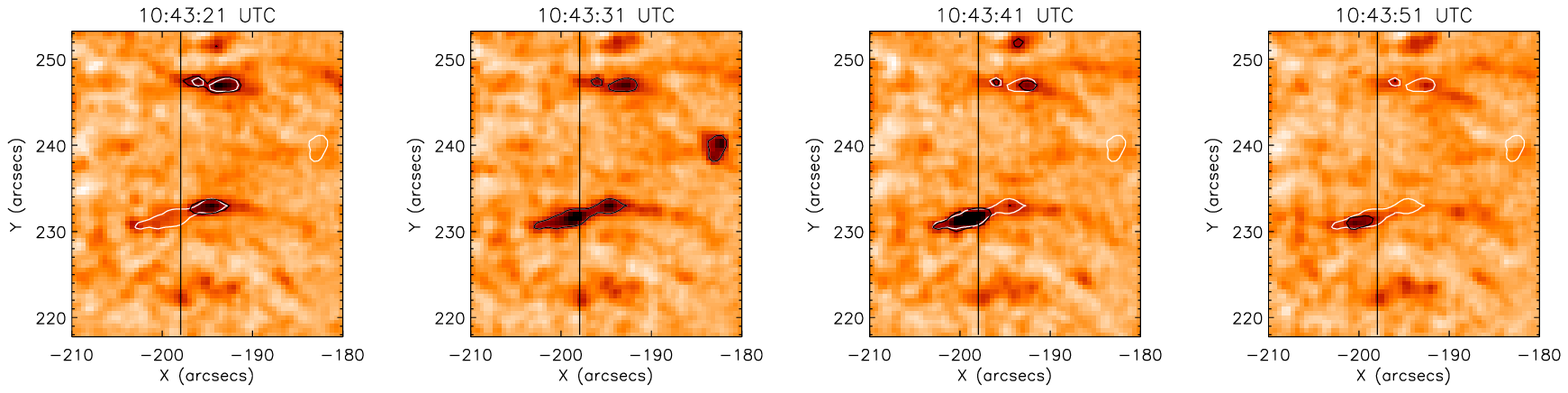}
	 \includegraphics{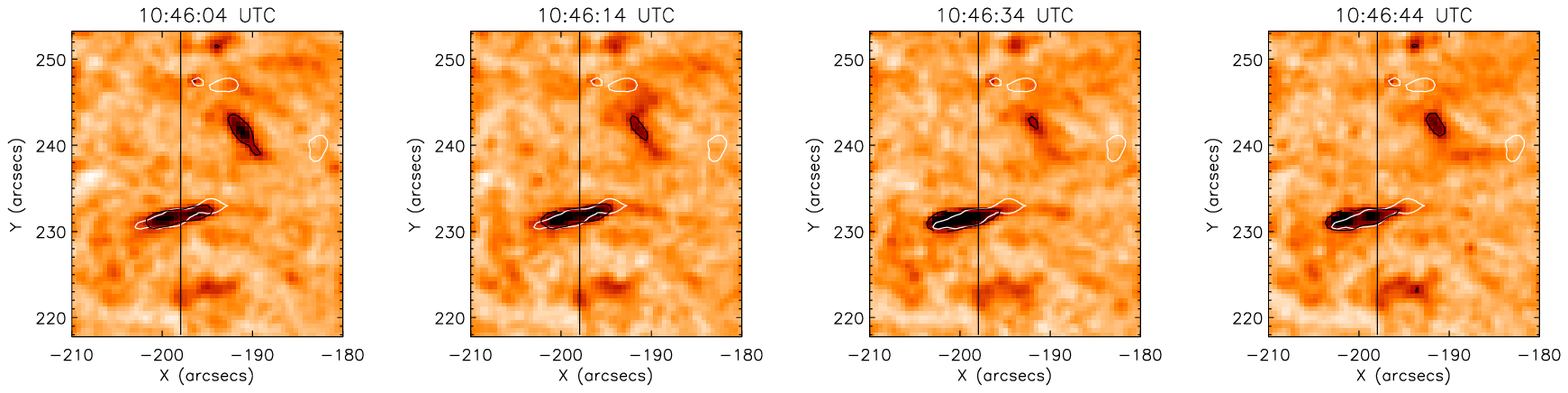}
	
	\caption{TRACE~$\lambda$171~\AA\ colour table reversed image sequences of events (c)
	(above) and (d) (below). The black contour shows the brightening (here seen as the 
	darkest feature) in the loop at a given time. The over-plotted white 
	contour outlines the event at 10:43:31~UTC. The black contour traces the 
	 brightenings at the given time. The vertical
	 line corresponds to the position of the SUMER slit. 
	\label{fig3}}
\end{figure*}

The TRACE \citep{1999SoPh..187..229H} data  were obtained in the Fe~{\sc ix/x}~$\lambda$171 
and $\lambda$1600~\AA\ passbands starting at 09:00~UTC and finishing at 11:30~UTC on 1999 
June 2. The integration time was 2.9~{\rm s} for the $\lambda$171~\AA\ passband and 
0.3~{\rm s} for $\lambda$1600~\AA. The $\lambda$171~\AA\ channel cadence was 10~{\rm s} 
which increased to 15~{\rm s} when an image in the $\lambda$1600~\AA\ channel was taken. 
From 09:18:38~UTC until 09:32:28~UTC only observations in the $\lambda$1600~\AA\ channel 
were taken. The FOV of the images was 256\arcsec\ $\times$ 256\arcsec\ with a spatial 
resolution of 1\arcsec. The most recent work on the TRACE~$\lambda$171~\AA\ passband 
temperature response can be found in \citet{2006ApJS..164..202B}.
 
The MDI \citep{1995SoPh..162..129S} data were taken with a cadence of 1 {\rm min} in 
 high-resolution mode (pixel size 0.606\arcsec) during several hours.

The co-alignment of TRACE and SUMER observations was done by using  TRACE~$\lambda$1600~\AA\ 
images and SUMER raster observations, taken just before the time series in the 
Si~{\sc ii}~$\lambda$1260.44~\AA\ (logT$_{max}$/K = 4.1) line which  falls in the 
transmitted O~{\sc v}~$\lambda$629.73~\AA\ spectral window.  The line formation 
temperatures are taken from CHIANTI v5.2 using the \citet{1998A&AS..133..403M} ionisation 
equilibrium. The SUMER raster was obtained 
with 5~{\rm s} exposure time and 0.37\arcsec\ increment. The emission in the 
TRACE~$\lambda$1600~\AA\ passband  mainly comes from continuum emission, C~{\sc iv}, C~{\sc i}, 
and Fe~{\sc ii}. Note that the SUMER times mark the beginning of the exposures, while 
the TRACE times the end of the exposures. The co-alignment between SUMER  and MDI was 
done via the TRACE~$\lambda$1600~\AA\ channel. The precision of all co-alignments is 
1\arcsec.

\begin{figure*}[ht!]
\centering
        \includegraphics[scale=0.8]{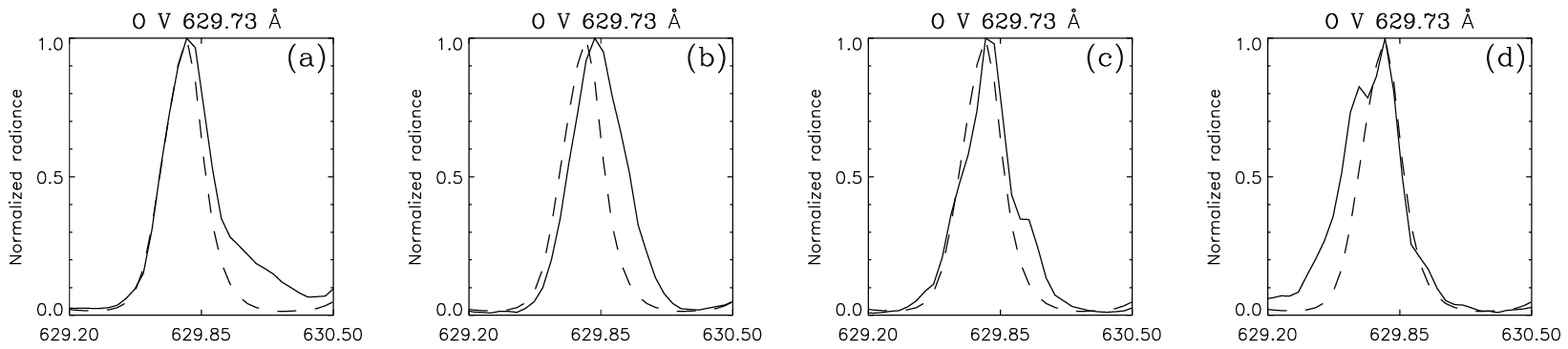}
	\includegraphics[scale=0.8]{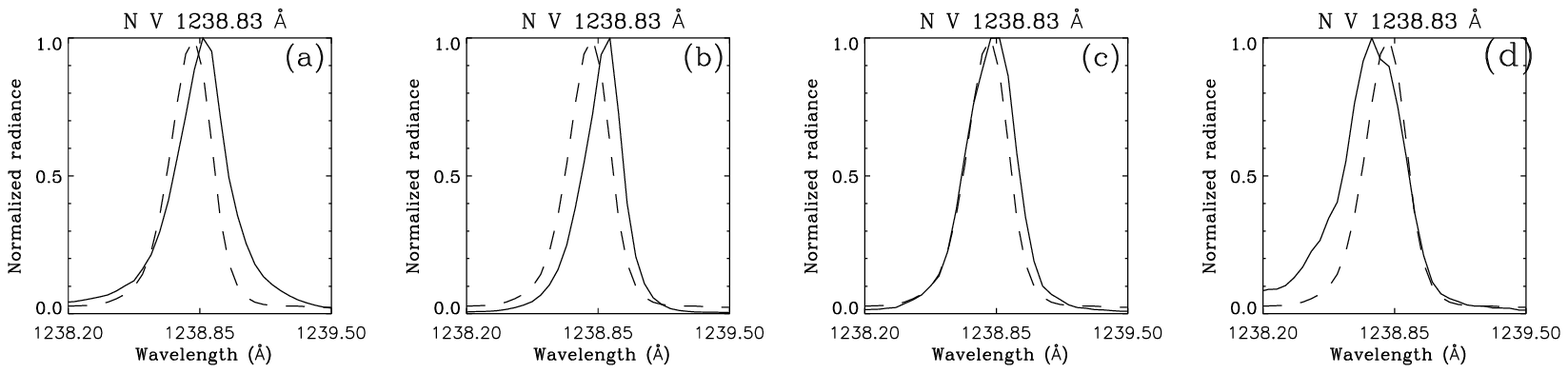}
	
	\caption{Normalised O~{\sc v} and N~{\sc v} line profiles of the four events shown 
	in Fig~\ref{fig2}. The dashed line is the reference spectrum.
	\label{fig4}}
	\end{figure*}

\section{Analysis and discussion of small-scale flows}

The SUMER  observations were taken in spectral lines with formation temperatures 
covering a temperature range from 10\,000~K to 1~MK. First, the two 
transition region lines O~{\sc v}~$\lambda$629~\AA\ and N~{\sc v}~$\lambda$1238~\AA\ were 
inspected for spectra showing non-Gaussian profiles or large intensity 
increases. That led to the identification of several events with profiles 
showing a radiance increase and either  blue or/and red-shifted emission. 
 The counterpart of these events in the TRACE~$\lambda$171~\AA\ images revealed several 
features of increased emission  of circular shape with diameters of 3\arcsec\ and 
elliptical ones with major axis of 4\arcsec\ and minor axis of 2\arcsec\  
along small-scale loops (length up to 30 Mm) (Fig.~\ref{fig2}). This combined with 
the spectral information mentioned above led us to conclude that we actually 
observe plasma flows. The brightenings as seen in the TRACE images appear 
suddenly anywhere along the loops. We can only speculate that either the energy 
deposition happens at the position where first these brightenings are seen or 
simply  due to higher density or filling factor that the plasma
emits stronger in the TRACE~$\lambda$171~\AA\ channel at this place. We selected 
four events which were best registered with the two instruments both 
temporally and spatially. The transient flows have a different response in the 
registered  spectral lines but nevertheless some common characteristics 
can be found. 

Events (a) and (b) presented in Fig.~\ref{fig2} (top) both show a strong 
increase in the peak radiance of the rest component of N~{\sc v} of 3.5 
and 12.5 times, respectively. The increase in the O~{\sc v} line is only a factor of 1.5 
and 3.4. This is coupled with an average increase of  2.2 and 3.8  times in the 
chromospheric lines (C~{\sc i}, Si~{\sc ii} and  S~{\sc ii}). How can we explain 
such a significant difference in the intensity flux increase of the two transition 
region lines O~{\sc v} and N~{\sc v}? One possible explanation is that although 
both lines have overlapping formation temperatures, their peak formation 
temperature is separated by $\Delta$logT$_e$ = 0.08--0.12 (depending on the assumed atomic model).
 A time dependent ionisation  could also be an explanation which will be 
discussed in a forthcoming paper. Another possible reason was discussed by 
\citet{2005A&A...439.1183D} where the authors derived the density dependent contribution 
function for both N~{\sc v} and O~{\sc v}. Their calculations showed that with 
increasing electron density, both lines shift towards lower temperatures. The 
difference comes, however, from the relative increase of the line flux with 
increasing density. For the N~{v} line, increasing the density to 10$^{11}$~cm$^{-3}$ 
results in a 60\% increase in the line flux, while the O~{v} line shows a 30\% decrease. 
Increasing the density to 10$^{12}$~cm$^{-3}$ results in a factor of two decrease of the O~{v} flux.

Event (a) shows a Doppler shift of 80--90 $\kms$  in the red wing of the O~{\sc v} line and 
a rest component shifted at  $\approx$ 5~$\kms$. For event (b) the rest 
component shift to the red in the O~{\sc v} line is $\approx$14 $\kms$ while the 
shift in the red wing of the line  reaches up to 60 $\kms$. The Doppler shifts 
in O~{\sc v} were derived from a multi-Gauss fit.

The flows (c) and (d) cross the SUMER slit with a time difference of approximately 3~{\rm min}.
 Using image difference techniques we found that these flows appear visibly at the same 
location, i.e. in the same loop (Fig.~\ref{fig2}, bottom and Fig.~\ref{fig3}). Visual 
inspection of the TRACE image sequence  animation suggests that the plasma crossing the 
 SUMER slit propagates in the same direction during both events. Despite moving along 
 the same magnetic loop and in the same direction, there is quite a significant difference 
in the spectral line profiles during the two events. The Doppler shifts in the O~{\sc v} 
line (only for O~{\sc v} was it possible to apply a multi-Gauss fit) change from red 
(24~$\kms$) and blue-shift (50~$\kms$) for event (c) to a  blue-shift of up to 45~$\kms$ 
with a strong peak at 28 $\kms$ and a red-shifted but not very intense component  up to 
38~$\kms$ for event (d) (Figs.~\ref{fig3} \& \ref{fig4}). What causes such a difference in 
the line profiles? One possible explanation is the line-of-sight effect created by the 
change of the orientation of the loop (due to, for instance, a footpoint displacement) 
in respect to the observer (SoHO/SUMER). We can only speculate that for case (c) 
the slit is closer to the top of the loop while for (d) the slit gets the 
emission more from the ascending plasma which produces the blue-shifted emission. 
As can be seen in Fig.~\ref{fig3}, the projection of the flow on the TRACE 
images differs slightly for events (c) and (d). 
 In fact, for these events, flows seems to propagate towards both footpoints 
of the loop when viewed with the TRACE images. 
The SUMER slit gets the emission only from one of them. A similar picture was 
also observed during other events registered in the TRACE field-of-view. However,  
we have to be cautious with the interpretation of what we see in 
the TRACE data, because without simultaneous spectral information we
 cannot be sure whether we observe a flow or simply a brightening.

Events (c) and (d) show a similar radiance increase which is stronger 
in  the O~{\sc v} line  (7.4 and 6.3 times with respect to the reference radiance, 
respectively) while the emission in N~{\sc v} increases only 4.6 and 4.3 times, respectively. 
This is coupled with a weaker response in the chromospheric lines with respect to events 
(a) and (b). It is therefore evident that a stronger response in the N~{\sc v} line is 
coupled with a stronger emission in the chromospheric lines (C~{\sc i}, 
Si~{\sc ii} \& S~{\sc ii}). 

\begin{figure}[h!]
\centering
        \includegraphics[scale=0.7]{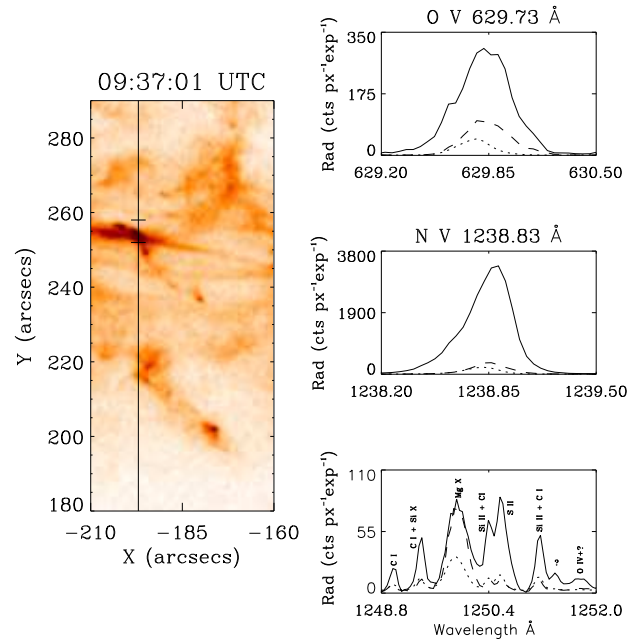}
	\caption{{\bf Left:} TRACE~$\lambda$171~\AA\ colour table reversed image showing 
	the flow in a pre-existing coronal loop. The vertical line corresponds to the 
	position of the SUMER slit, while the two horizontal lines outline the part of the slit
        which was analysed in the SUMER data. {\bf Right:} A reference spectrum is 
	shown with the dotted line. The dashed line corresponds to the spectrum registered 
	at 09:22:17~UTC averaged over the area denoted by the two horizontal lines on 
	the TRACE image. The continuous line is the spectrum around 09:37 UTC.}
	\label{fig5}
\end{figure}


\begin{figure}[h!]
\centering
        \includegraphics[scale=0.8]{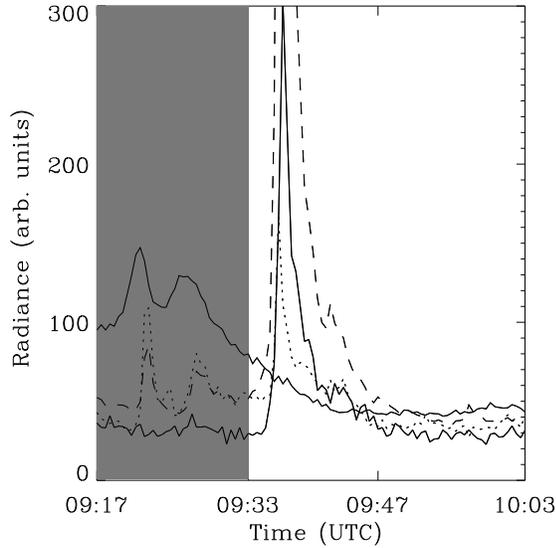}
	\caption{ The peak radiance averaged along 
	the SUMER slit as shown in Fig.~\ref{fig5} obtained from a single Gauss fit. 
	The solid line corresponds to the Mg~{\sc x}~$\lambda$624.85~\AA\ line, dotted -- 
	O~{\sc v}~$\lambda$629~\AA, dashed -- N~{\sc v}~$\lambda$1238~\AA, solid thick -- 
	Si~{\sc ii}~$\lambda$1251.16~\AA\ and  C~{\sc i}~$\lambda$1251.17~\AA. The grey 
	area marks the time period during which the jet-like events were observed 
	\citep[see][]{2007ApJ...670L..57M}.}
	\label{fig6}
\end{figure}

Although there are some significant differences in the radiance increase of O~{\sc v} 
and N~{\sc v}, the Doppler shifts in the two transition region lines are almost the same 
for each transient (Fig.~\ref{fig4}). If we assume that O~{\sc v}~$\lambda$629~\AA\ 
and N~{\sc v}~$\lambda$1238~\AA\ are in ionisation equilibrium they will emit at 
temperatures around  250\,000 and  200\,000~K which gives  sound 
speeds of 75~$\kms$ and 67~$\kms$, respectively.  The fact that the flows direction
forms an angle with respect to the line-of-sight means  that some of these flows 
reach supersonic velocities.

The interpretation of the Mg~{\sc x}~$\lambda$624.85~\AA\ emission is complicated due 
to the fact that in the SUMER  spectra this line when registered in second order is 
 blended by several other lines including P~{\sc ii}~$\lambda$1249.82~\AA, 
Mg~{\sc ii}~$\lambda$1249.93 \AA, Si~{\sc ii}~$\lambda$1250.08 \AA\ and 
O~{\sc iv}~$\lambda$1250.25 \AA\ in second order. Events (b), (c) and 
(d) show a significant increase in the feature at $\lambda$1249.90~\AA\ but 
this is coupled with the presence of the O~{\sc iv} multiplet observed in second order 
(see Table~1 for details on these lines).  Therefore, it remains uncertain 
whether  these events reach coronal temperature, but equally we cannot reject 
such a possibility. Additionally, during some of the events,  
two spectral lines appear at around  $\lambda$1250.39 \AA\ and $\lambda$1250.78~\AA\ 
blended with the O~{\sc iv}~$\lambda$1251.70 \AA\  indicated with a question mark on the spectra 
in Fig.\ref{fig2}. A possible candidate for the line at $\lambda$1251.39 is Si~{\sc v}. 
This line has a very high excitation energy of the upper level, and therefore it can be 
excited only if the electron distribution function has an enhanced energy tail around 
logT/K= 5.2 -- 5.5 (P.~R. Young, private communication). The second line at around 
$\lambda$1250.78~\AA\ is not yet identified. Further work is needed regarding the 
identification of these lines. 

The duration of the flows was determined from the time a  brightening is  first seen in the loop
 in the TRACE~$\lambda$171~\AA\ image until it disappears. The events last from 60~{\rm s} 
 in a very small loop (event (b)) up to 19~{\rm min} (event (a)) for the largest loop. 
 The duration in the SUMER data is based on the time the flow is seen crossing the slit 
 and does not actually reflect the real lifetime of the events.

\section{Line-of-sight effects revealed in imager and spectrometer  co-observations}

One of the aims of the present work is to show how a line-of-sight effect can give 
misleading information on the observed features when either only an imager or a 
spectrometer is used. In Fig.~\ref{fig5} a TRACE~$\lambda$171~\AA\  image illustrates a 
jet-like feature \citep[for more details on this event see][]{2007ApJ...670L..57M}  
which is `crossed' by a blob-like flow along a loop (similar to the events discussed above). 
 With imager information only, it would not be clear whether the two apparently crossing 
features are related to each other or not, and what their plasma characteristics are.
 Equally, the SUMER line profiles 
show a strong increase in all spectral lines  in the transmitted spectral windows.
 As can be seen from Table~1 the spectral lines  in question cover a temperature range 
 from 10\,000~K to 1~MK. The fact that we have both imaging and spectral information
helps us to disentangle this puzzle. To find out the origin of the emission 
in each spectral line we took a spectrum obtained just before the blob
flow was seen in the TRACE images. The corresponding profile at 09:22:17~UTC is 
shown with a dashed line in Fig.~\ref{fig5}. This profile has only
information from the jet-like feature. The comparison  with the reference spectrum 
(dotted line) and the spectrum (solid line) at 09:37:01~UTC (during the blob-like flow)
shows that the emission in the Mg~{\sc x} line comes entirely from the jet-like
event. A small amount of the O~{\sc v} emission belongs to the jet-like event 
(this is actually the fainting phase of the phenomenon, see Fig.~\ref{fig6})  
while the remaining emission belongs to the blob-like flow. The jet-like event as shown by 
\citet{2007ApJ...670L..57M} emits only at coronal and
transition region temperatures and no chromospheric emission was detected during
the event(s). Therefore, the emission from the chromospheric lines belongs entirely 
to the blob-like flow. So, what we actually see in the TRACE~171~\AA\ channel 
are two features happening at very different heights in the solar atmosphere which are 
not related to each other. 

\section{Conclusions}

This paper presents a unique dataset which makes it possible to 
study features in the plage area of an active region using imager and spectrometer 
co-observations at the highest existing temporal, spectral (in the wavelength in question) 
and spatial resolution. We show that Doppler shifts often associated with
 explosive events or bi-directional jets can actually be identified with flows 
 (some of them reaching supersonic velocities) in small-scale loops.  We were able 
 to determine the physical properties of these flows such as temperature, velocities,
 size and lifetime.  The present results can add, however,
 very little on  the understanding of the actual mechanism generating these flows, but 
 they can have a strong application for testing different models of generating plasma 
 flows from any energy source like for instance magnetic reconnection. The present work 
 will continue using data from Extreme-ultraviolet Imaging Spectrometer, X-ray Telescope 
 and Solar Optical Telescope on board Hinode combined with data from already well  known
 instruments such as SUMER, Coronal Diagnostic Spectrometer, MDI on SoHO and TRACE. 
  Large datasets of simultaneous observations with the above mentioned instruments were 
 acquired and preliminary work is ongoing. Hinode data should provide the answer  to
 some of the open questions of the present work like for instance the coronal response
 to these transient phenomena.

We  believe that with this work we were able to demonstrate that using simultaneous 
imager and spectrometer observations is of great importance for deriving correctly 
the plasma characteristics of the observed features  and that this is one of the best ways 
to better understand the physical processes happening on the Sun.

\begin{acknowledgements}
 We would like to thank W. Curdt for the fruitful discussions. We thank A. Theissen 
 and D. E. Innes for the careful reading of the manuscript. SoHO is 
 a mission of international co-operation between ESA and NASA. The SUMER project is
 financially supported by DRL, CNES, NASA, and PRODEX. Research at Armagh 
 Observatory is grant-aided by the Northern Ireland Department 
 of Culture, Arts and Leisure.

\end{acknowledgements}

\bibliographystyle{aa}

\end{document}